\documentclass[showpacs]{revtex4}
\usepackage{graphicx}
\usepackage{bm}

\begin{document}
\title{Effect of General Relativity and rotation on the energy 
deposition rate for 
$\nu + \overline{\nu} \rightarrow e^+ + e^- $ inside a compact star }
\author{Abhijit Bhattacharyya}
\email{abphy@caluniv.ac.in}
\affiliation{Department of Physics, University of Calcutta, 92, A. P. C. Road;
Kolkata - 700009; INDIA}
\author{Sanjay K. Ghosh}
\email{sanjay@bosemain.boseinst.ac.in}
 \affiliation{Centre for Astroparticle Physics and
Space Science \& Department of Physics; Bose Institute; 93/1, A.P.C Road;
 Kolkata - 700009; INDIA}
\author{Ritam Mallick}
\email{ritam@bosemain.boseinst.ac.in}
 \affiliation{Centre for Astroparticle Physics and
Space Science \& Department of Physics; Bose Institute; 93/1, A.P.C Road;
 Kolkata - 700009; INDIA}
\author{Sibaji Raha}
\email{sibaji@bosemain.boseinst.ac.in}
 \affiliation{Centre for Astroparticle Physics and
Space Science \& Department of Physics; Bose Institute; 93/1, A.P.C Road;
 Kolkata - 700009; INDIA}
\pacs{26.60.+c}
\begin{abstract}
We have studied the $\nu + \overline{\nu} \rightarrow e^+ + e^- $ energy 
deposition rate in a rotating compact star. This reaction is important 
for the study of gamma ray bursts. 
The General Relativistic (GR) 
effects on the energy deposition rate have been incorporated. We find that the 
efficiency of the process is larger for a rotating star.
The total energy deposition 
rate increases by more than an order of magnitude due to rotation. The 
dependence of this energy deposition 
rate on the deformation parameter of the star has also been discussed. 

\end{abstract} 
\maketitle

\section{Introduction}
Gamma Ray Burst (GRB) and its possible connection with neutrino production in  
compact stars is a field of high current interest. GRBs were first discovered 
in the late 1960s by U.S. military satellites \cite{key-1}.
Until recently, GRBs were perhaps the biggest mystery in high energy
astronomy. To unravel this mystery, several satellite based detectors have
been employed for the observation of GRB. 
GRBs are separated in two classes; long duration bursts (long GRB)
which last from 2 sec. to several minutes, with average duration
of 30 seconds and
short duration bursts (short GRB) with burst duration from few
milliseconds to 2 seconds, average being 0.3 seconds \cite{key-2}.
                                                                                
Initial evidence that the long GRBs are associated with supernovae came
from the study of GRB 980425 in 1998 \cite{key-3}. This burst was 
tentatively linked
to supernova SN 1998bw.  The definitive proof came on March 29, 2003, when
a relatively nearby burst, GRB 030329, produced an afterglow whose optical
spectrum was nearly identical to a supernova \cite{key-4}.
In contrast, there is a scarcity of information on short GRBs. Until recently, 
information was only available about the burst; the post-burst picture
was not clear. Of late, the detection of afterglows
in short GRBs and also precise localizations of different short GRBs have
provided some more inputs for the study of these phenomena.

On the otherhand, compact stars (neutron or quark stars) are objects formed 
in the aftermath of supernova. The central density of these stars can be as 
high as $10$ times that of normal nuclear matter. Due to beta equilibration,
a large number of neutrinos and antineutrinos may be produced inside the 
compact stars. These neutrinos and antineutrinos could annihilate and 
give rise to electron-positron pairs through the reaction $\nu {\bar \nu} 
\rightarrow e^+ e^-$. These $e^+ e^-$ pairs may further give rise to gamma 
rays which could be a possible explanation of the observed GRBs. Hence it 
is very important to study the energy deposition in the $\nu {\bar \nu}$ 
annihilation process. 

Previous calculation of this reaction in the vicinity of a neutron star has 
been based on newtonian gravity  \cite{key-20,key-21}, {\it i.e.}  
$( 2GM/c^2R ) << 1 $, where $M$ is the gravitational mass of the neutron star 
and $R$ is the distance scale. The effect of gravity was incorporated in 
refs. \cite{key-22,key-23}, but only for a static star. In our present 
calculation, we extend the basic premise to rotating stars. 

In this work first we will discuss the equation of state (EOS), the metric 
and the structure of the star. The effect of GR and also that of rotation 
will be discussed next and finally we will present our results. 

\begin{figure}[h]
\vskip 0.4in
\centering
\includegraphics{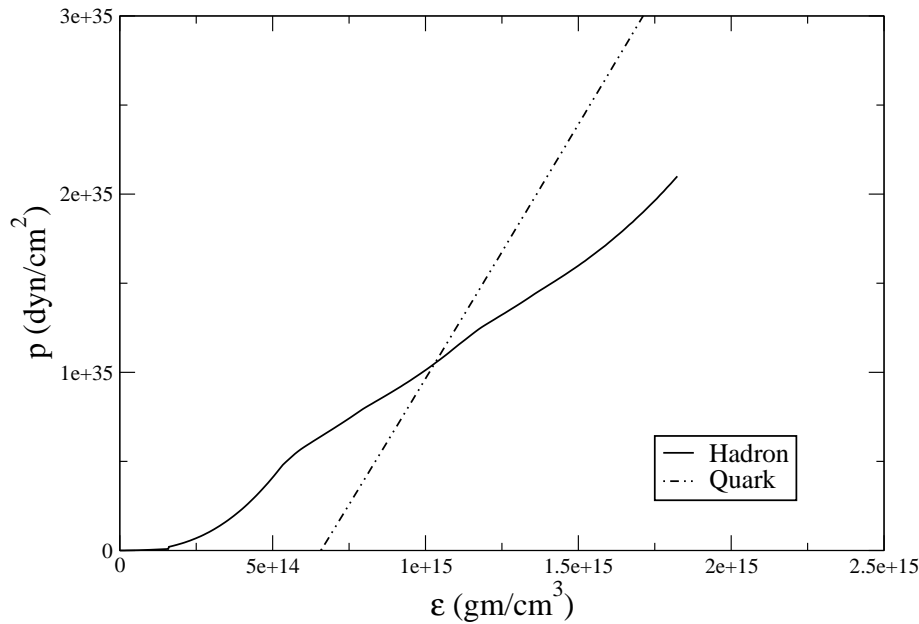}   
\caption{Variation of pressure with energy is plotted for both hadronic and 
quark EOS.}
\end{figure}

\section{Star Structure}
The structure of the star is described by the metric given by \cite{key-24}
\begin{eqnarray}
ds^2 = -e^{\gamma+\rho}dt^2 + e^{2\alpha}(dr^2+r^2d\theta^2) + 
e^{\gamma-\rho}r^2 sin^2\theta(d\phi-\omega dt)^2
\end{eqnarray}
The four gravitational potentials, namely $\alpha, \gamma, \rho$ and 
$\omega$ are functions of $\theta$ and $r$ only. Once these potentials are 
known, the observed properties of the star can be evaluted. 
The Einstein's equations for the three potentials 
$\gamma, \rho$ and $\omega$ have been solved using
Green's function technique \cite{key-25,key-26}. The fourth potential 
$\alpha$ has been determined from other potentials. All the potentials 
have been solved for both static as well as rotating stars using the 
{\bf 'rns'} code; the details of the code may be obtained in ref. \cite{key-27}. 

The solution of the Einstein's equations needs an EOS as an input. In the present work 
we have used a non-linear version of the Walecka model with TM1 \cite{key-27a} parameter
set. For comparison, we have also used a quark matter EOS obtained from standard 
Bag model with $B^{1/4} = 160 MeV$. These EOSs have been plotted 
in figure 1. The quark matter EOS is more stable at higher energy 
densities. Using these EOSs, we can get 
the structure of the star by solving the Einstein's equations. The solution of 
Einstein's equations also provide the density profiles of the star. 
These profiles, for a star rotating in the mass shedding limit i.e. with 
the Keplerian velocity, are plotted in fig. 2 and fig. 3. 
The neutrino annihilation process has been studied for both neutron and quark stars
with the central energy density $1.2\times10^{15} gm/cm^{3}$. 
Fig. 2 shows the variation of energy density with the radius of the star 
for $\chi = cos\theta=0$, {\it i.e.} along the equator. Here $\theta$ is the 
angle that radius vector makes with the polar axis. From fig 2. we can see that
the energy density is high at the centre of the star, and as we move 
radially outwards the energy density decreases to a lower value. Fig. 3 shows 
the variation of the energy density with $\chi$. This is plotted for radius 
$r=3.5 Km$ from the centre of the star. The 
curve shows that for $\chi =0$, {\it i.e.} at the equator, the energy density 
is maximum and it falls off gradually as we approach the pole $\chi=1$. We 
would like to mention at this stage that for the hadronic matter we have 
considered a thin crust. On the other hand, quark matter being self bound,
no such crust has been used as for quark star. Hence the energy density at the 
surface of a quark star is much higher compared to that of a hadronic 
star (fig. 2).

\begin{figure}
\vskip 0.4in
   \centering
\includegraphics{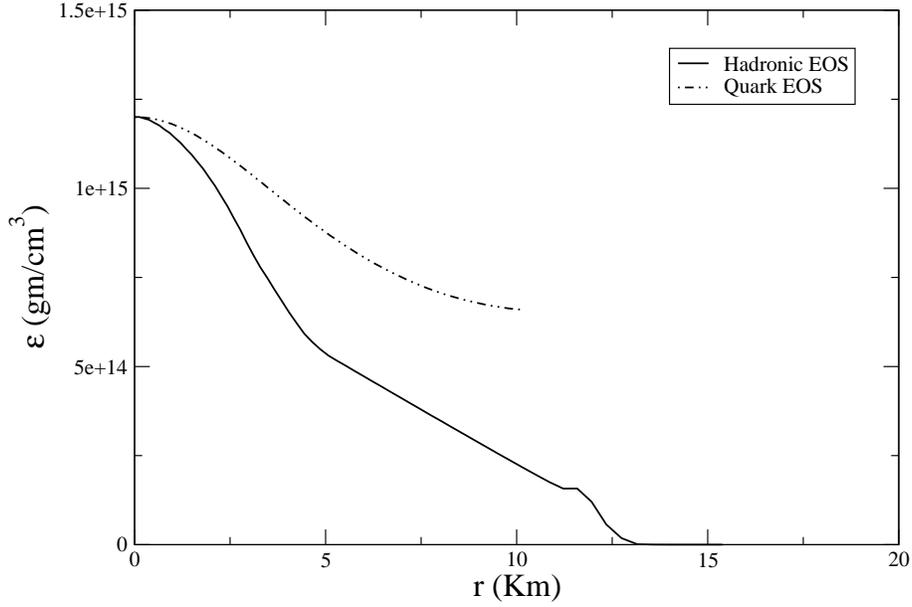}   
\caption{Variation of energy density along the equator of the star which 
is rotating with the Keplerian velocity.} 
\end{figure}

\begin{figure}
\vskip 0.4in
   \centering
\includegraphics{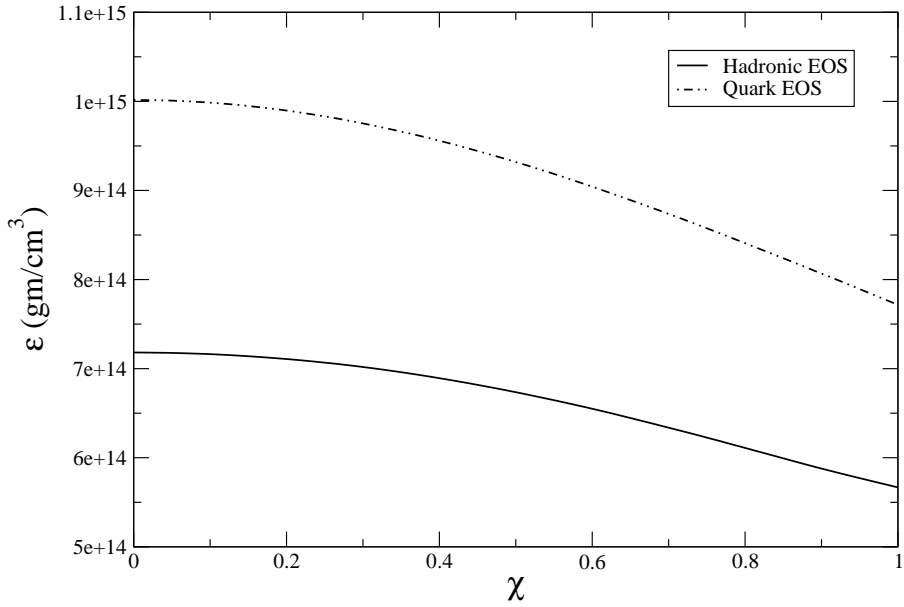}   
\caption{Variation of energy density with $\chi$ at distance of $3.5 Km$ from 
the centre of the star.}
\end{figure}

\section{GR effects}
As already mentioned above the energy 
deposition rate due to the process $\nu {\bar \nu} \rightarrow e^+ e^-$ in 
Newtonian gravity  has been studied earlier \cite{key-22,key-23}. 
The energy deposited per unit volume 
per unit time in Newtonian gravity can be given as \cite{key-21}

\begin{eqnarray}
\dot{q}(r)= \int \int f_{\nu}(p_{\nu},r) f_{\overline{\nu}}(p_{\overline{\nu}},
r) [\sigma|v_{\nu}-v_{\overline{\nu}}|\epsilon_{\nu} 
\epsilon_{\overline{\nu}}]\frac{\epsilon_{\nu} + \epsilon_{\overline{\nu}}}
{\epsilon_{\nu} \epsilon_{\overline{\nu}}} d^3 p_{\nu} d^3 p_{\overline{\nu}},
\label{qdot}
\end{eqnarray}
where $f_{\nu}$ and $f_{\overline{\nu}}$ are the number densities of neutrinos 
and antineutrinos respectively in the phase space, $v_{\nu}$ is the neutrino 
velocity, and $\sigma$ is the rest frame cross section for the process 
$\nu {\bar \nu} \rightarrow e^+ e^-$. In eqn. \ref{qdot}, we have, 
\begin{eqnarray}
[\sigma|v_{\nu}-v_{\overline{\nu}}|\epsilon_{\nu} 
\epsilon_{\overline{\nu}}] = \frac{DG_f^2}{3\pi} 
(\epsilon_{\nu} \epsilon_{\overline{\nu}}- p_{\nu}.p_{\overline{\nu}})^2 .
\end{eqnarray}
The above expression is Lorentz invariant. ${G_f}^2 = 
136 \times 10^{-24} MeV^{-4}$, and $D = 1 \pm 4sin^2 \theta_W +8sin^4 
\theta_W$. Here $sin^2 \theta_W=0.23$. We will consider the plus sign only 
in the expression of $D$ since this corresponds to the $\nu_e {\bar \nu}_e$ 
pairs. It has been assumed that the mass of the electron is negligible. Now we 
replace $p_{\nu} = \epsilon_{\nu} \Omega_{\nu}$ and 
$d^3 p_{\nu}=\epsilon_{\nu} d\epsilon_{\nu} d\Omega_{\nu}$, 
where $\Omega_{\nu}$ is the unit direction vector and $d\Omega_{\nu}$ is the 
solid angle. So the rate of energy deposition is 
\begin{eqnarray}
\dot{q}(r)= \frac{DG_f^2}{3\pi } \Theta (r) \int \int f_{\nu}
f_{\overline{\nu}}(\epsilon_\nu + \epsilon_{\overline{\nu}})
\epsilon_\nu^3 \epsilon_{\overline{\nu}}^3 d\epsilon_{\nu} 
d\epsilon_{\overline{\nu}} 
\end{eqnarray}
where the angular integration is represented by
\begin{eqnarray}
\Theta (r) = \int \int (1- \Omega_{\nu}.\Omega_{\overline{\nu}})^2 
d\Omega_{\nu} d\Omega_{\overline{\nu}}.
\end{eqnarray}
So the energy and angular dependences can be decoupled to make the 
evaluation simpler.

Let us now consider the effects of GR on this process. The effect of GR 
on static star was studied by Salmonson and Wilson \cite{key-22}. 
In the present paper, we consider the static as well as the rotating star. 
The GR effect will modify both the 
energy and angular integrals and will provide us with a new 
lower limit of integration.

Let us first consider the path of a zero mass particle, 
{\it i.e.} a null geodesic. For the metric considered here, one gets, 
\begin{eqnarray}
(\frac{1}{r^2e^{\gamma-\rho}}\frac{dr}{d\phi})^2=\frac{1}{b^2}
\frac{1}{e^{2\alpha}
(e^{\gamma+\rho}-e^{\gamma-\rho}r^2\omega^2)}+\frac{1}{b}\frac{2\omega}
{(e^{\gamma+\rho}-e^{\gamma-\rho}r^2\omega^2)}-\frac{1}{e^{2\alpha}}
\frac{1}{r^2e^{\alpha-\rho}},
\end{eqnarray}
where $r$ is the distance from the origin, $\phi$ is the latitude and 
$b$ is the impact parameter. From the above equation, one can immediately 
see that the geodesic explicitly depends on the gravitational potentials 
{\it i.e.} on the EOS and the frequency of rotation. If we now follow the 
technique used by \cite{key-22} and express this equation in 
terms of the angle $\theta$ between the particle trajectory and the tangent 
vector to the circular orbit, we get
\begin{eqnarray}
(\frac{dr}{d\phi})^2=tan^2\theta\frac{e^{\gamma-\rho}r^2}{e^{2\alpha}}.
\end{eqnarray}
If the above expression is substituted in eqn. (6), we get a quadratic 
expression for $b$.
\begin{eqnarray}
b^2-\frac{2\omega r^2}{e^{2\rho}-r^2\omega^2}b-\frac{r^2}{e^{2\rho}-r^2
\omega^2}=0.
\end{eqnarray}
This equation can be solved to get
\begin{eqnarray}
b=\frac{\omega r^2\pm re^{\rho}}{e^{2\rho}-r^2\omega^2}.
\end{eqnarray}
b will be same for all the points lying on the same orbit. The 
above equation also implies a minimum photosphere radius, which we would 
denote by $R$, below which a massless particle (neutrino) emitted tangentially 
to the stellar surface ($\theta_R =0$) would be gravitationally bound.

To carry out the angular integral in eqn. (5), we define $\lambda = sin \theta$,
and express the solid angle \( \Omega \) as
\begin{eqnarray}
\Omega=[\lambda,(1-\lambda^2)^{\frac{1}{2}} cos\phi,(1-\lambda^2)^{\frac{1}{2}} sin\phi]\\
d\Omega=cos\theta d\theta d\phi.
\end{eqnarray}
Using the same notation, we further obtain
\begin{eqnarray}
\Theta (r) = 4\pi^2 {\int_x}^1 {\int_x}^1 [1-2\lambda_{\nu}.\lambda_{\overline{\nu}}+{\lambda_{\nu}}^2.{\lambda_{\overline{\nu}}}^2+\frac{1}{2}(1-{\lambda_{\nu}}^2)(1-{\lambda_{\overline{\nu}}}^2)]d\lambda_{\nu}.d\lambda_{\overline{\nu}}.
\end{eqnarray}
where $\lambda_\nu$ and $\lambda_{\bar \nu}$ are for neutrinos 
and antineutrinos 
respectively, and 
\begin{equation}
x(r)=\sqrt{1-\left(\frac{R}{r}\right)^2 \frac{e^{2\rho}-r^2\omega^2}
{e^{2\rho_R}-R^2\omega_R^2}}
\end{equation}
$\rho_R$ being the potential $\rho$ at the photosphere. Carrying out
the integration over $\lambda_{\nu}$ and $\lambda_{\overline{\nu}}$, we get
\begin{eqnarray}
\Theta(r)=\frac{2\pi^2}{3}(1-x(r))^4(x(r)^2+4x(r)+5).
\end{eqnarray}
The gravitational effect modifies the temperature in the energy integral. The 
temperature $T(r)$ in the energy integral eqn. (4) is the neutrino 
temperature at radius $r$. The temperature of the free streaming neutrinos at 
radius $r$ in terms of their temperature at the minimum photosphere radius 
$R$ can be deduced if we assume that the temperature varies linearly with 
redshift. Therefore
\begin{eqnarray}
T(r)=\sqrt{\frac{e^{2\rho_R}-R^2\omega_R^2}{e^{2\rho}-r^2\omega^2}}T(R).
\label{temp}
\end{eqnarray}
In order to quantify the total $e^+ e^-$ pair energy deposition, we define 
${Q}$ as the integral of $\dot{q}$ over proper volume. ${Q}$ is a 
measure of the total amount of energy converted from the neutrinos to 
$e^+ e^-$ pairs over the whole volume per unit time. 
So the total amount of local energy deposited 
via $\nu + \overline{\nu} \rightarrow e^+ + e^- $ reaction is given by 
\begin{eqnarray}
Q=2\pi {\int_R}^{surface} dr \int  \dot{q}(r) r^2 sin\theta 
\frac{e^{2\alpha +(\gamma-\rho)/2}}{\sqrt{1-v^2}}d\theta .
\end{eqnarray}
with $v=(\Omega-\omega)r sin\theta e^{-\rho}$, $\Omega$ being the rotational 
velocity of the star and $\frac{1}{\sqrt{1-v^2}}$ is the lorentz factor.

Integration over the radial and angular variable gives us the total 
energy deposited per unit time by the $\nu {\bar \nu}$ pairs. 

\begin{figure}[h]
\vskip 0.4in
\centering
\includegraphics{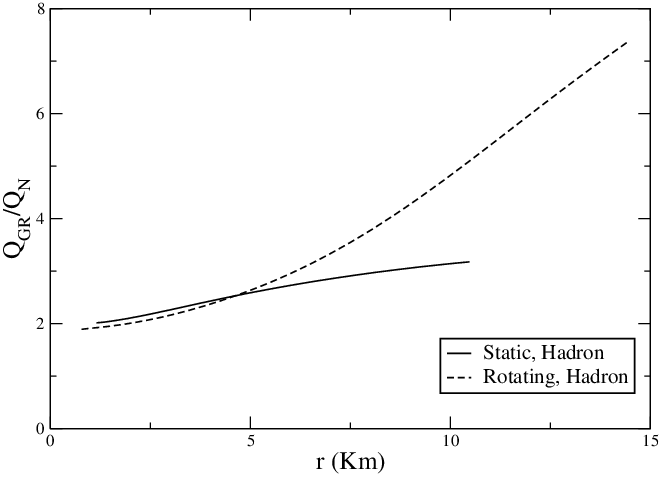}   
\caption{Variation of the ratio of rate of energy deposition for general 
relativistic (GR) over Newtonian is plotted along the radial direction, 
for the hadronic star.}
\end{figure}

\section{Results}
The temperature at the minimum photosphere $T(R)$ is kept at a 
small value of 5 MeV to evaluate the integral for $Q$. In the present study,
we have assumed that there is no neutrino trapping. It has been shown earlier
\cite{skgnpa} that neutrino trapping may increase the temperature considerably
inside the star. The eqn. (\ref{temp}) shows that the 
temperature of the star decreases as we move radially outwards because 
temperature, like energy, varies linearly with redshift. For the static star, 
we find that $b$ is minimum for $r=R=1 Km$ for the hadronic star and 
$r=R=1.2 Km$ for quark star. For a rotating star due to 
high rotational velocity of the star, the shape of the star becomes oblate 
spheroid. The star is therefore no longer spherically symmetrical and it has 
an extra deformation parameter to describe its shape. We define this 
deformation parameter $\chi=cos\theta$, along the vertical axis of the star. 
For the EOS, which we have taken the {\bf 'rns'} code gives the rotational 
velocity of the star to be $\Omega=0.62 \times 10^4s^{-1}$ for the hadronic star 
and $\Omega=1.4 \times 10^4s^{-1}$ for the quark star, the central energy 
density of both the stars being $1.2 \times 10^{15} gm/cm^3$. 
For such a configuration the star deforms to a oblate spheroid but the
 photosphere is found to be spherical. Photosphere is evaluated by
minimising $b$ for different $\chi$. The photosphere radius comes out to be 
$r=R=0.78$ Km for the hadronic 
star and $r=R=0.6$Km for the quark star.

\begin{figure}[t]
\vskip 0.4in
\centering
\includegraphics{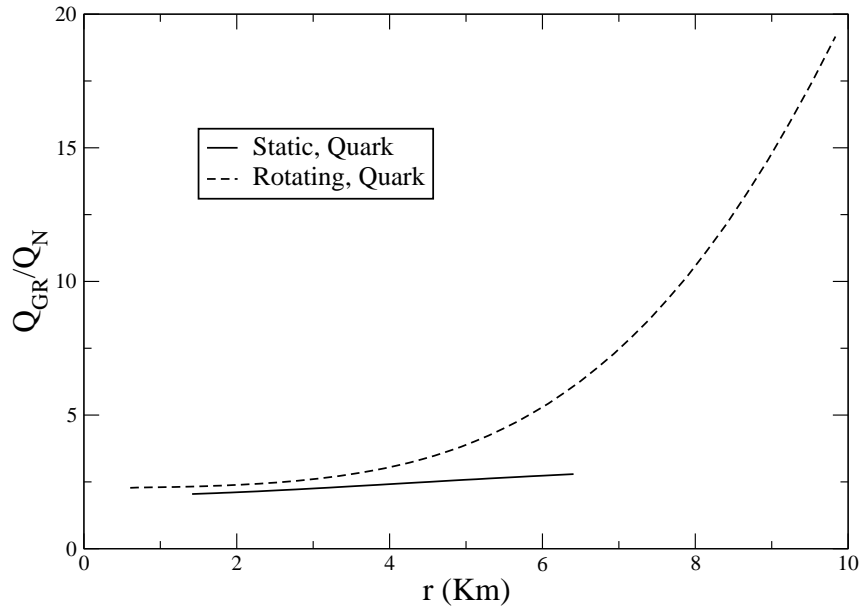}   
\caption{Variation of the ratio of rate of energy deposition for general 
relativistic (GR) over Newtonian is plotted along the radial direction, 
for the quark star.}
\end{figure}

\par
Figures 4 and 5 depicts the importance of GR on the energy deposition rate both 
for the static and the rotating compact stars. In both these figures we have divided the 
radius of the star into small bins of length $100m$ and integrated $Q$ over 
those bins. The curves in fig. 4 show that for a static star, the energy 
deposition increases by a factor of $2$ at the photosphere.  As we go towards 
the surface the deposition increases and at the surface of the star this 
ratio becomes little over $3$. For the rotating star, at the surface, 
this ratio is a little less than $8$. For the hadronic star the two curves 
intersect. The ratio for the quark star is about $4$ for the static case 
and reaches a much higher value of $20$ at the surface when the star is 
rotating. From these two figures we conclude that the effect of rotation 
is to enhance the energy deposition by an order of magnitude both for the 
quark star and the hadronic star. Moreover, a comparison of the figures 4 and 5 
also gives the effect EOS on the energy deposition rate.
Since most of the other results are qualitatively similar, in the rest 
of the paper we will use hadronic star for further discussions.  

\begin{figure}[h]
\vskip 0.3in
   \centering
\includegraphics{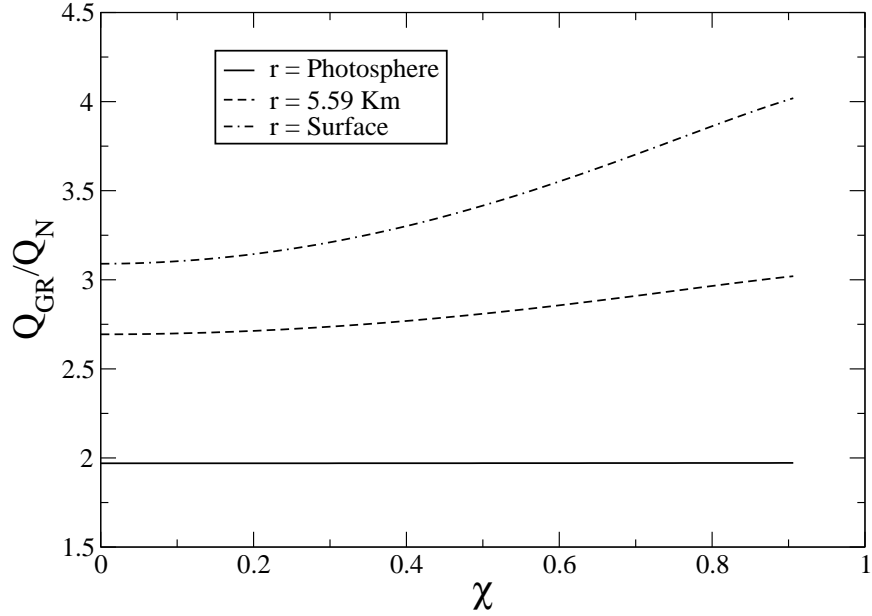}
\caption{Variation of the ratio of rate of energy deposition with $\chi$,
 for three different radial points inside the neutron star.}
\end{figure}

\begin{figure}[h]
\vskip 0.4in
   \centering
\includegraphics{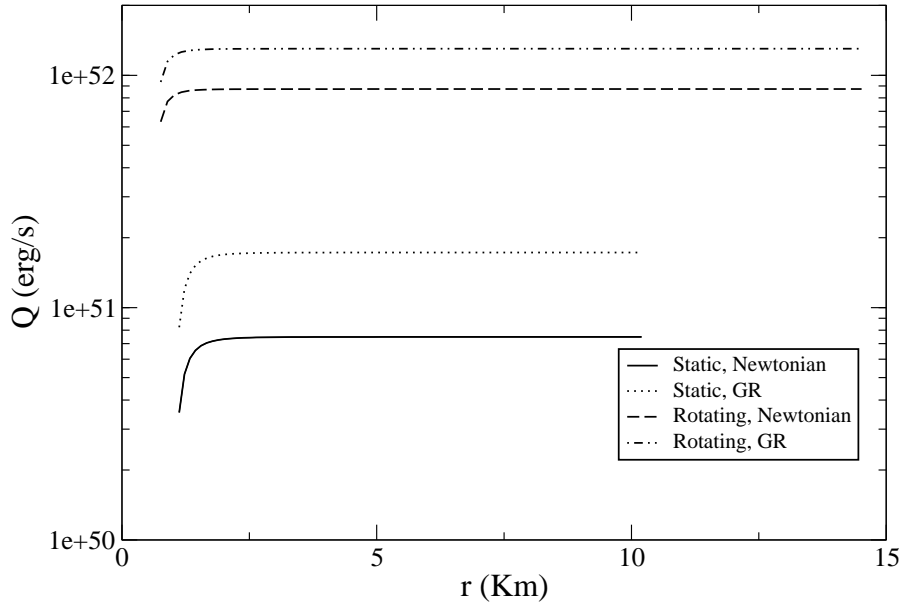}
\caption{Variation of rate of energy deposition of general 
relativistic (GR) and Newtonian along the radius of the static 
and rotating neutron stars.}
\end{figure}

\par
The figures 6 and 7 show the effect of
 GR and rotation on the energy deposition rate.
In fig. 6 the variation of ${Q_{GR} \over Q_{N}}$ has been plotted as a 
function of $\chi$, for different $r$.  Here, one can see that the 
ratio changes with the polar angle due to 
rotation. The deposition rate is somewhat pronounced at the pole compared 
to the equator, mainly near the surface of the star.
 In fig. 7 we have plotted the total energy deposition rate along the radial 
direction. The rate of total energy deposition is maximum at the photosphere 
and it decreases radially outwards, which is implied by the 
fact that the curves saturate to a maximum limit at higher radius. 
From this figure one can see that due to rotation the total energy deposition 
increases by an order of magnitude to $10^{52} ergs/s$.
It should be mentioned at this stage that the stars (both static and rotating)
are constructed with equal central density ($1.2 \times 10^{15} gm/cm^3$). The 
gravitational masses for static and rotating stars are $1.57 M_\odot$ and 
$1.94 M_\odot$ respectively for the hadronic EOS and $1.13 M_\odot$ and 
$2M_\odot$ respectively for quark matter EOS. We have also studied the 
energy deposition rates for constant mass sequences. The qualitative 
features remain the same and the quantitative results change only by a few 
percent. For comparison we should mention that Salmonson and Wilson 
\cite{key-22,key-23} did the 
GR calculation for such a reaction on a static star using the Schwarzschild 
metric. They found that for type II supernova models the enhancement is by a 
factor of $2$ at a rather larger radius $R=10M$, increasing to a factor of 
$4$ at $R=5M$, where M being the total mass of the star. For a collapsing 
neutron star, the more extreme relativistic regime, the enhancement is by a 
factor of $30$. Recent numerical 
calculation of neutron stars in closed binary system has been done in 
ref. \cite{key-28,key-29}. It has been argued that $0.5\times 10^{53} ergs$ 
may be liberated from a star in thermal neutrinos within a few 
seconds \cite{key-30}.

To conclude, we have studied the effect of GR and rotation on the energy 
deposition rate of the reaction $\nu {\bar \nu} \rightarrow e^+ e^-$. 
The effect of rotation on such a reaction has been considered for the 
first time. Our findings show that these effects enhance the energy deposition 
rate by more than an order of magnitude. This finding is very important in the 
context of describing the GRB associated with a compact, massive, rotating object. 
In our present calculation we have not considered any neutrino trapping. 
However the neutrinos may be trapped and raise the temperature of 
the star \cite{skgnpa,key-19}, thereby further changing the energy deposition rate. 
Furthermore the consideration of non-equilabrated quark matter rather than 
equilibrated matter may enhance the energy deposition further. Such studies 
are in progress.

\acknowledgements{ R.M. would like to thank CSIR; New Delhi,
 for financial support. A.B. would like to thank CSIR; New Delhi,  
for financial support through the project 03(1074)/06/EMR-II.  S.K.G. 
and S.R thank DST; govt of India,  for financial support under the IRHPA scheme.}


\begin{thebibliography}{99} 
\bibitem{key-1} R. Klebesadel, I. Strong and R. Olsen, Astrophys. J Lett.
                182 (1973) L85; 
                E. P. Mazets, E. P. Golenetskii and V. M. Ilinskii, 
                JETP Lett. 19 (1974) 77
\bibitem{key-2} C. Kouvelioton {\it et. al.}, Astrophys. J Lett. 413 (1993) L101
\bibitem{key-3} T. Galama {\it et. al.}, Astrophys. J. 536 (2000) 185
\bibitem{key-4} J. Hjorth {\it et. al,}, Nature 423 (2003) 847; 
                T. Matheson {\it et. al.}, Astrophys. J. 599 (2003) 394;  
                K. Z. Starck {\it et. al.}, Astrophys. J. 591 (2003) L17
\bibitem{key-20} J. Goodman, A. Dar and S. Nussinov, Astrophys. J. 314 (1987) L7
\bibitem{key-21} J. Cooperstein, L. J. van der Horn and E. Baron, Astrophys. J. 309 (1986) 653
\bibitem{key-22} J. D. Salmonson and J. R. Wilson, Astrophys. J. 517 (1999) 859
\bibitem{key-23} J. D. Salmonson and J. R. Wilson, Astrophys. J. 578 (2002) 310
\bibitem{key-24} G. B. Cook, S. L. Shapiro and S. A. Teukolsky, Astrophys. J. 422 (1994) 227 
\bibitem{key-25} H. Komatsu, Y. Eriguchi and I. Hachisu, Mon. Not. R. Astron. Soc. 237 (1989) 355 
\bibitem{key-26} A. Bhattacharyya, S. K. Ghosh, 
 M. Hanauske and S. Raha, Phys. Rev. C 71 (2005) 048801
\bibitem{key-27} N. Stergioulas and J. L. Friedman, Astrophys. J.
 444 (1994) 306
\bibitem{key-27a} J. Ellis, J. I. Kapusta and K. A. Olive, Nucl. Phys. B 348 (1991) 345; 
Y. Sugahara, H. Toki, Nucl. Phys. A 579 (1994) 557
\bibitem{skgnpa} S. K. Ghosh, S.C. Phatak and P.K. Sahu, Nucl. Phys. A 596 (1996) 670
\bibitem{key-28} G. J. Mathews and J. R. Wilson, Astrophys. J. 482 (1997) 929
\bibitem{key-29} G. J. Mathews, P. Marronetti and J. R. Wilson, Phys. Rev. D 58 (1998) 043003
\bibitem{key-30} J. R. Wilson, J. D. Salmonson and G. J. Mathews, AIP Conf. Proc. 428 (1997)
 Gamma-ray Bursts: Fourth Huntsville Symposium, ed. C. A. Meegan et al. (New York:AIP), 788
\bibitem{key-19} A. Bhattacharyya, S. K. Ghosh 
 and S. Raha, Phys. Lett. B 635 (2006) 195 
%
\end{thebibliography}
\end{document}